# A Novel Combined Term Suggestion Service for Domain-Specific Digital Libraries

Daniel Hienert, Philipp Schaer, Johann Schaible and Philipp Mayr

GESIS – Leibniz Institute for the Social Sciences,
Lennéstr. 30, 53113 Bonn, Germany
{Daniel.Hienert, Philipp.Schaer, Johann.Schaible, Philipp.Mayr}@gesis.org

**Abstract.** Interactive query expansion can assist users during their query formulation process. We conducted a user study with over 4,000 unique visitors and four different design approaches for a search term suggestion service. As a basis for our evaluation we have implemented services which use three different vocabularies: (1) user search terms, (2) terms from a terminology service and (3) thesaurus terms. Additionally, we have created a new combined service which utilizes thesaurus term and terms from a domain-specific search term recommender. Our results show that the thesaurus-based method clearly is used more often compared to the other single-method implementations. We interpret this as a strong indicator that term suggestion mechanisms should be domain-specific to be close to the user terminology. Our novel combined approach which interconnects a thesaurus service with additional statistical relations outperformed all other implementations. All our observations show that domain-specific vocabulary can support the user in finding alternative concepts and formulating queries.

## 1 Introduction

A general and long known problem with keyword-based search is the so called "vocabulary problem" or "wording problem" [6]. The same information need or search query can be expressed in a variety of ways. Current web search engines often retrieve a list of documents where same relevant items are always included – but this is mostly a phenomenon of the very large document index. Thus, when using a "wrong" term there is still a high probability getting a non-empty result set.

When we analyze today's Digital Library (DL) systems or domain-specific databases a controlled vocabulary, usually a thesaurus is used to index the publications. DLs often consist of metadata entries on the specific publications, descriptive abstracts are optional. In this situation the vocabulary problem can become quite serious. If the searcher doesn't use one of the controlled terms the document was indexed, the chance of getting relevant documents is low. There is a significantly higher chance to retrieve an empty result set. Users tend to adapt their search strategies to work around these drawbacks. In a user study done by Aula et al. [1] one expert arti-culated: "I choose search terms based not specifically on the information I want, but rather on how I could imagine someone wording […] that information."

Modern information-seeking support systems (ISSS) try to make use of a variety of automated approaches to transform and expand textual queries e.g. by using stop word lists, stemming or spelling correction. From the perspective of interface design interactive query reformulation still is an open research issue. [11].

In the following paper we will present the results of a user study with more than 4,000 unique visitors in the online information portal Sowiport[1]. Users were confronted with three basic term suggestions services based on (1) user-search-terms, (2) terms from a terminology service and (3) terms from a domain-specific thesaurus. As a novel approach, we have created a term suggestion service that combines thesaurus terms and terms from a domain-specific search term recommender. We will present related work in section 2, followed by the evaluated vocabularies and services in section 3. We will proceed with the conducted evaluation in section 4 and will present results in section 5. We conclude this paper with a discussion in section 6.

## 2 Related Work

We will present two different perspectives on query reformulation tools: the origin of the proposed terms and the different types of reformulation tools.

As Efthimiadis [5] points out interactive query expansion (IQE) can be divided in two types of IQE mechanisms: (1) those that are based on collection dependent or independent knowledge structures and (2) those that are based on the search results. The difference between these approaches is the origin of the data to propose terms from. The terms that are presented to the user can either be retrieved from a knowledge structure like e.g. thesauri or from the documents that are included in the search result (e.g. to perform a pseudo-relevance feedback). Regarding this characteristics Vechtomova et al. [18] compared two approaches for query expansion (QE) based on term co-occurrences. The first approach was a global co-location analysis where the entire document collection was used to extract related terms. The second approach only used terms from a local subset of the retrieved documents. This local approach clearly performed better than the global one. The difficulties in the first global approach seemed to lie in proposing too unspecific and too general terms [3]. The authors argued that users need to have a more context specific QE mechanism.

The fact that users need supporting mechanisms to correctly formulate their queries is supported by the user studies of Hargittai [7]. She found that 63% of the participants made a typographical or spelling mistake of some kind, and among these, 35% made only one mistake, but 17% made four or more errors during their entire session. This is supported by the search engine logs analysis of Cucerzan and Brill [4]. They found that 10 – 15% of the queries had typographical error.

Hearst [8] described two types of supporting: spelling suggestions/corrections and automated term suggestions. Term suggestion can be further differentiated in pure term and query suggestion like shown by Kelly et al. [10]. While a term suggestion is only focused on single terms, a query suggestion tries to combine suggested terms with other terms and present them to the user as a new and complete query. Query suggestions therefore can provide alternative viewpoint and can help to explore unfamiliar scientific areas. Like Kelly showed users preferred the query suggestion method and rated it higher. This included the ability to help them think about new approaches for their search.

---

[1] http://www.gesis.org/sowiport

Query and term suggestions are implemented in many modern web search engines but many systems only try to make suggestions based on prefix matches (user types "soc" and the system suggests "social", "society" and so on) while the actual origin of the suggestions remains unclear. A typical representation of this kind of suggestion was shown by White and Marchionini [20]. They performed a study on an interactive method, which they called "real time query expansion". After the user types a word and presses the space bar, the system presents terms based on the surrogates of the ten top-ranked documents. On these prototypes White et al. [19] conducted a usability study with 36 participants, each doing two known-item tasks and two exploratory tasks, and each using the baseline system, the query suggestions, and two other experimental interfaces. For the known-item tasks, the query suggestions scored better than the baseline on all measures ("easy", "restful", "interesting", etc.). Participants of their study were also faster using the query suggestions over the baseline on known item tasks and made use of the query suggestions 35.7% of the time. Those who preferred the query suggestion interface said it was "useful for saving typing effort" and "for coming up with new suggestions".

A more general study on Web search interfaces was performed by Jansen et al. [9]. They studied a search engine log file from Dogpile.com with 2.5M interactions (1.5M of which were queries) from 2005. Using their computed session boundaries (mean length of 2.31 queries per session), they found that more than 46% of users modified their queries, 37% of all queries were parts of reformulations, and 29.4% of sessions contained three or more queries.

Regarding the combination of these approaches Schatz et al. [16] did a study on two different term suggestion methods: one using terms from a subject thesauri and the other from term co-occurrence lists. The overall finding was that multiple approaches resulted in a better search quality. They suggest combining different IQE methods and origins in favor of a single method.

In the following sections we will describe our different term suggestion services and the set-up of our evaluation.

## 3   Term Suggestion Services

We implemented three basic term suggestion services with different vocabularies as a basis for our evaluation: user-search-terms (UST), a terminology service (HTS) and a social science thesaurus (TS). Additionally, as a novel approach, we have created a combined term suggestion service (CTS) which combines the social science thesaurus and a search term recommender service.

### 3.1 Basic Term Suggestion Services

We use different vocabularies as a data basis for the basic term suggestion services in our study that are introduced in the next sections: (1) User-Search-Terms (UST), (2) Terminology Service (HTS), and (3) Thesaurus Terms (TS).

The service UST is an uncontrolled set of terms extracted from the query log of the social science portal Sowiport. We recorded about 28,000 distinct terms entered by human users since 2007. The applied service includes all user terms from the query log as a flat list of terms without any additional information (see Fig. 1a). The terms are chosen by matching the input term against a ranked list of the user terms. The user terms are ordered by the frequency count of their usage.

The service HTS is a controlled set of terms coming from a terminology service (called heterogeneity service) implemented in the portal Sowiport [12]. The service contains controlled terms from 25 different thesauri with about 26,500 distinct terms. The thesauri are connected with intellectually created relations that determine equivalence, hierarchy (i.e. broader or narrower terms), and association mappings between terms. To search and retrieve terminology data from the heterogeneity service an individual has to enter correct terms from at least one controlled vocabulary of the service. For the recommendation service we used an adapted heterogeneity service that returns a flat list of related terms ordered alphabetically (see Fig. 1b).

The Thesaurus for the Social Sciences (Thesaurus Sozialwissenschaften) is an instrument to index and retrieve subject-specific information in Sowiport. The list of keywords contains about 11,600 entries, of which more than 7,750 are descriptors and about 3,850 are non-descriptors. Topics in all of the social science disciplines are included. The applied service includes all descriptors from the thesaurus as a flat list of terms without any relational data and ordered alphabetically (see Fig. 1c).

### 3.2 Combined Term Suggestion Service

As a data basis for our Combined Term Suggestion Service we use thesaurus terms and terms from a Search Term Recommender (STR). Petras [14] proposed a search term suggestion system, which relies on two basic parameters: (1) the controlled vocabulary terms that are used for document representation and (2) the natural language keywords that are input by the searcher. The advantage of suggesting controlled vocabulary terms as search terms is that these terms have been systematically assigned to the documents, so that there is a high probability of relevant and precise retrieval results if these terms are used instead of whatever natural language keywords the searcher happens to think of.

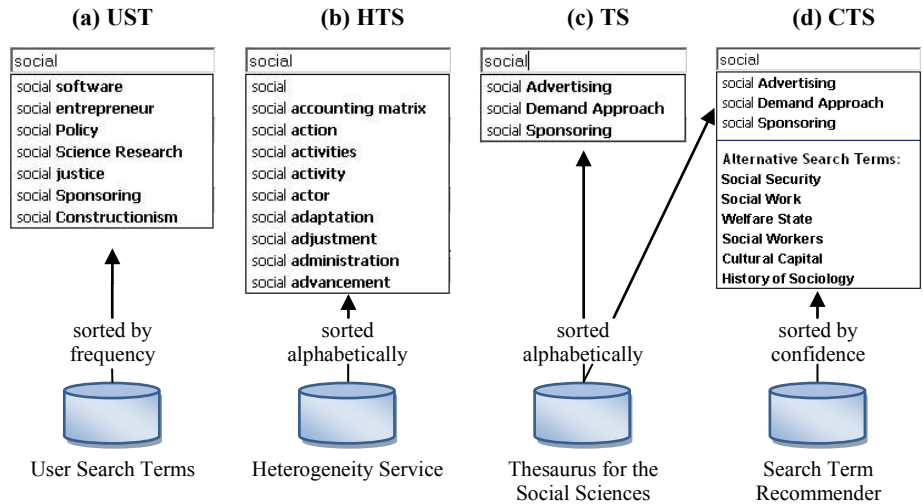

**Fig. 1.** Four implemented term suggestion services: (a) User-Search-Terms (UST), (b) intellectually mapped terms (HTS), (c) Social-Science Thesaurus (TS) and (d) Combined Term Suggestion (CTS).

The STR addresses the problem of search term vagueness by performing a co-word analysis of the terms of a field in order to recommend more appropriate terms to the user. The STR maps query terms to indexing terms at search time by building term-term-associations between two vocabularies: natural language terms from titles and abstracts on the one hand side and controlled vocabulary on the other hand side. The associations are weighted according to their co-occurrence within the collection to predict which of the controlled vocabulary terms best mirror the search terms.

In the original implementation Plaunt and Norgard [15] used a likelihood ratio statistic to measure the association between the natural language terms from the collection and the controlled vocabulary terms to predict which of the controlled vocabulary terms best mirror the topic represented by the searcher's search terms. Given a training set of documents containing free terms form title/abstract and controlled vocabulary terms a dictionary of co-words can be build which includes strength of association (the calculated weight). This can be used to predict the possibility of likelihood between words.

Our own implementations rely on latent semantic analysis and support vector machines. They are applied via the commercial indexing software Mindserver. The used service returns a flat list of terms corresponding to the input term ordered by the strength of association.

The combined term suggestion service combines the TS service with recommendations from the search term recommendation service. Until three characters it shows terms from the thesaurus, beginning with four characters it shows an additional section with *Alternative Search Terms* under the TS list. We used the limit of four letters to avoid input terms for the STR that leads to poor results. Term suggestions that appear in both lists are filtered out and are shown only in the TS section.

## 4 Evaluation

In this section we first present the social sciences information portal Sowiport as a real-world environment for our user study. In section 4.2 we describe the logging process and the evaluation periods.

### 4.1 Evaluation Environment

We chose the social science information portal Sowiport as a real-world environment for our user study. Sowiport integrates literature references, persons, institutions, projects, services and studies. It currently contains about 4.8 million literature references and research projects from 18 databases, including six databases from ProQuest/CSA, which are available by a national license funded by the German Research Foundation. The German-language share of the databases include the GESIS own databases SOLIS and SOFIS, which contain about 500,000 literature references and research projects which are indexed intellectually with the Thesaurus for the Social Sciences. Sowiport is offered in German and English, the majority of users are from German-speaking countries. The portal reaches about 7,000 unique visitors per month. The term suggestion functionality has been integrated in the simple search form on the home page and in the advanced search form. The term suggestions are proposed to the user as a list under the input field. The user can choose a term from the list with a mouse click or by scrolling and return. The term is then entered into the input field. With a click on the button *Search* or with *Enter* the search is submitted.

### 4.2 Logging Process and Evaluation Periods

For conducting the user study, we had to log the entered search terms, the selection from the recommendation's list and the search queries performed. Taking this information from the server log can be a very time consuming and error-prone issue. Server logs are full of records from irrelevant search engines and crawlers and we want to make sure only to log data from human users. We therefore implemented a function that logs all this information only if a user clicks on the search button or hits enter in the search field. In particular we have logged the following information: (1) for a selection of a recommendation from the list: entered term, chosen term, position of the chosen term, service type, date/time and session id. (2) For a submitted search: submitted term, date/time and session id.

The different vocabularies UTS, HTS, TS and CTS were sequentially activated in Sowiport in a time period of about 3 months. Each service was activated until the count of visitors using the search had exactly reached 1000 unique visitors. Once the number was reached we changed the service to the next one. A unique visitor is identified technically by an internal ID. The user can perform several actions like browsing or searching in the database, but is recognized only once as a unique user. A user session is still valid for two hours if the user performs no further actions.

## 5  Results

In this section we will show the individual results of the conducted user study. In section 5.1 we will present results of the use of term suggestion services and in section 5.2 we will show a categorization of patterns we have found in the data.

### 5.1  Use of Term Suggestions Services

We used the unit measure of 1000 unique visitors to calculate the share of selected recommendations to all users. This number describes the average use of search term suggestions based on unique users. For the CTS approach 50.9% of the users used the recommendation service, followed by the TS vocabulary with 37.5%, the UST vocabulary with 25.2% and the HTS with 10.4%.

As a second measure we have calculated the share of selected recommendations to all searches performed. This number describes the use of recommendations based on all searches and shows therefore the general use of search term suggestions. The CTS performed best with 14% usage, then the TS vocabulary with 9%, the UST with about 7% and the HTS service with only about 3% usage. The number has been in all cases under 15%, which means, at best, only in one of seven search queries the recommendation service has been used, in the case of the HTS service only in three of one hundred searches. In general we can say that there is a very weak use of the recommendation service, one would expect a much higher usage. Possible reasons, such as the count of letters entered to the choice of term, the word length of the chosen term etc. are discussed in the following paragraphs.

On the basis of the collected data we were able to calculate the average position of the selected term from the recommendations list and the average count of letters entered to the choice of the term. On average the selected term had the second position on the list. Figure 2 shows the distribution of the position of the chosen term in the list. The graphs of the individual vocabularies and services show a similar trend. The percentage of selection of the concept at first to tenth position decreases, this means, recommendations on a high position in the list are chosen clearly more often.

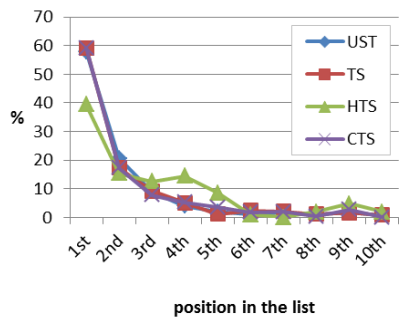 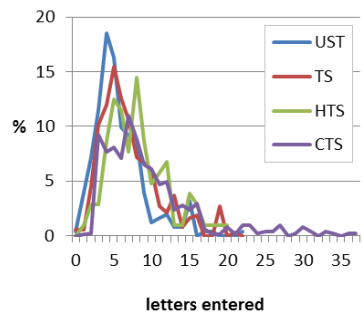

**Fig. 2.** Rank of the chosen term in the list of recommendations.

**Fig. 3.** Number of letters entered until a term is chosen.

The user enters about nine letters, before he chooses a term. The average word length of the chosen concept has been 15 letters, including terms and term combinations. This is more than twice the average word length of German words of 6.44 [13]. This means the user selects very long terms and term combinations. Even with excluding terms combinations, the average word length of individual chosen terms with about 14 letters is still very high. The distribution of the word length of the entered term in figure 3 shows peaks from 6 to 11 letters. To summarize, this leads to the following conclusions:

- Recommendations were used at best for every seventh search query.
- Recommendations on a high position in the list are chosen clearly more often.
- For terms with short and normal word length recommendations are not selected.
- In contrast, terms and term combinations with very long word length are selected.

**Table 1.** Summarized results: key figures for each technique, their use in relation to unique users and search queries and measures for entered and chosen terms.

|  | UST | HTS | TS | CTS |
|---|---|---|---|---|
| Unique users | 1000 | 1000 | 1000 | 1000 |
| Search queries | 3566 | 3572 | 4165 | 3604 |
| Selected recommendations | 252 | 104 | 375 | 509 |
| Share of selected recommendations to all searches | 7.06% | 2.91% | 9% | 14.12% |
| Share of selected recommendations to all unique users | 25.2% | 10.4% | 37.5% | 50.9% |
| Average position of the selected concept | 2 | 2.9 | 2.1 | 2.1 |
| Average count of letters entered to the choice of the term | 6.8 | 9.2 | 7.7 | 11 |
| Average word length of chosen concepts (single terms and combinations) | 15.1 | 16.6 | 14.8 | 15.2 |
| Average word length of single terms | 13.2 | 15.6 | 13.6 | 13.6 |

### 5.2 Patterns of Use

The data indicates different patterns of use, which can be classified in different categories. We analyzed the evaluation data, identified different categories for the transition from entered to chosen term and classified them into these categories. We found four different categories for all services, two more categories for the User-Search-Terms vocabulary and one category for statistically near terms of the CTS approach. The different categories are:

1. *Simple term completion*: the user enters initial letters and chooses a concept from the list.
2. *Selecting an already completely entered term*: the user enters a concept completely and then selects the same concept from the list.
3. *Selecting an already completely entered term, after a simple term completion in the search before*: in the first search the user enters initial letters and chooses a concept from the list, in the following search the user enters the complete concept and then chooses the same concept from the list.
4. *Term extension*: the user enters a complete term and then chooses a concept with term extension from the list.
5. *Two complete terms entered, second one changed*: the users enters a concept with two terms completely and then chooses a concept from the list where the second term changes.
6. *Selecting a more abstract concept*: the user enters a fine-grained concept and then chooses a more abstract concept from the list.
7. *Statistically near term*: the user enters a term and chooses a statistically near term form the *Alternative Search Terms*-section of the CTS approach. In contrast to the other categories, the terms have no common stem between entered and chosen term, but are statistically near.

**Table 2.** Examples for the different categories.

| Category | Entered term | Chosen term |
|---|---|---|
| 1 | acci | accident |
| 2 | accident | accident |
| 3 | acci<br>accident | accident<br>accident |
| 4 | accident | accident analysis |
| 5 | cognitive maps | cognitive development |
| 6 | mother-child clinic | mother |
| 7 | medicine | Doctor-patient-relationship |

**Table 3.** Frequency of categories in different services with more than 2%. For CTS individual results for each section and total results for the whole service are shown.

| Category | UST | HTS | TS | CTS | |
|---|---|---|---|---|---|
| 1 | 52.98% | 52.89% | 64.27% | 49.71% | 2.35% |
| | | | | 52.06% | |
| 2 | 9.92% | 16.34% | 13.6% | 13.75% | 0.4% |
| | | | | 14.15% | |
| 4 | 36.11% | 30.77% | 20.53% | 4.9% | 6.29% |
| | | | | 11.19% | |
| 7 | | | | | 22% |
| | | | | 22% | |

Simple term completion is used in more than 50% of cases (TS: 64.27%, HTS: 52.89%, CTS: 52.06%, UST: 51.98%) in all different services. The high ratio in the TS vocabulary indicates that here the proposed terms most likely correspond to the ones the user thought of. Selecting an already completely entered concept from the list is a pattern that occurs regularly: 16.64% in the HTS, 14.15% in the CTS, 13.6% in the TS and 9.92% for the UST vocabulary. The explicit selection from the list might be a cognitive assurance coupled with an action that chooses a concept in the sense of a controlled vocabulary. The user might think that if the system proposes a concept, it must be correct even if one enters the term himself before. The selection of an already entered term, after a simple term completion in the search before occurred six times only in the TS vocabulary. Here the user might have learned that (1) the concept exists and (2) that it leads to (useful) results. (1) is proven to the user through the appearance of the concept in the recommendations list, (2) is proven to the user by already seeing the result list for this search term. Term extension is the second big ratio for term completion: 36.11% for the UST vocabulary, 30.77% for the HTS, 20.53% for the TS and 11.20% for the CTS.

Categories 5 and 6 are patterns that occurred only within the User-Search-Term vocabulary. In category five the user enters a concept consisting of two terms and then chooses a concept from the list where the second term is changed, what happens three times. Selecting a more abstract concept has been a very rare occasion with only two examples: entered term: "mother-child clinic"/selected term: "mother" and entered term: "antidiscrimination eu"/selected term: "antidiscrimination".

Category 7 exist only for the CTS service and contains statistically near chosen terms, which are not simple term completions or extensions with common stem to the entered term, but are statically near terms based on co-word analysis. The number of 22% for the whole service therefore represents the number that could only be achieved by this particular service.

## 6 Discussion

In this study we evaluated an interactive term suggestion service for the domain-specific DL Sowiport. The service was tested with three different vocabularies (UST, HTS and TS) and a combination of TS and STR.

Term acceptance was generally comparable to other studies dealing with term suggestion methods, where the thesaurus-based method clearly scored best. Here the acceptance rate was between 37.5% and 50% (see table 1) which can be compared to other user studies in this field [19, 20]. The thesaurus-based method clearly outperformed the other single-method implementations which is a strong indicator that term suggestion mechanism need to be domain-specific to gain acceptance from the users.

This can be explained with the "Anomalous State of Knowledge" [2] wherein the user is while formulating queries. In this state he tries to map the words and concepts describing his problem to the terms of the system while typically fighting ambiguity and vagueness of language. This problem especially occurs in highly specialized scientific literature databases where often literature reference with spare bibliographic metadata is available for matching.

In scientific communities special discourse dialects evolve. These dialects are not necessarily the same dialects an information specialist or user would use to describe a document or a concept using a documentation language. The consequence is a serious source of vagueness in the query formulation phase. A term suggestion method per-se can support the user in this early stage of the search process e.g. choosing an appropriate query term which is used in the language of documentation.

It can be easily seen that different implementations of term suggestion services match different suggestion tasks: The simple term-completion tasks (category 1) are best matched by the TS, while near terms (category 7) are only suggested by the STR. UST can best match the need to extend terms with different concepts (category 4). While the categories 1 and 7 can be explained with their immanent features to be an expression of the language of indexation (in case of the TS) and an expression of the language of discourse (in case of the STR), the UST represents the language of the user looking for information. Here we can see the unfiltered search terms and queries users are actually using.

User studies in digital libraries have shown that most users are not aware of the special controlled vocabularies used in digital libraries [17]. Hence they are not using them in their query formulation. This can be seen in our relatively low acceptance rates of 9% of the thesaurus-based implementation. To overcome the acceptance problems we derived the plan to combine the rather simple term but most accepted completion method (TS) with a more sophisticated term-mapping (STR) approach. We can observe a significantly increase of the acceptance rate because these approaches complete each other.

On the one hand the TS service contains terms which are very relevant to the domain but on the other hand it is quite limited in its ability to deliver suggestions for all possible user aspects. This is due to its size of around 7,750 terms which have to map after three entered characters. In contrast the STR can nearly always suggest a controlled term since it maps a total of 1.45 million free terms on the 7,750 TS terms. The overall advantages of suggesting controlled vocabulary terms as search terms is that these terms have been systematically assigned to the documents, so that there is a high probability of relevant and precise retrieval results.

Our paper, especially the CTS approach, introduces new possibilities of suggestions in domain-specific DL. In the case of the STR we can see that term suggestions could provide a broader overview over different areas of a scientific domain or discussion, which typically involves particular associated concepts (perhaps assuming different meanings or directions of thought). The result is a diverse domain perspective on certain concepts, an effect that can also be achieved by displaying the semantic term mappings themselves. The general assumption of our paper is that term suggestions from several fields of research can provide a new view or different domain perspective on a topic in an interactive way. Combining different specific term suggestion methods is not an academic exercise; quite on the contrary, our approach has been clearly confirmed by users in a large scale real-life scenario.